\input harvmac
\sequentialequations
\def\Title#1#2#3{#3\hfill\break \vskip -0.35in
\rightline{#1}\ifx\answ\bigans\nopagenumbers\pageno0\vskip.2in
\else\pageno1\vskip.2in\fi \centerline{\titlefont #2}\vskip .1in}

\def\frac#1#2{{\textstyle{#1\over#2}}} 

\def\bra#1{\langle #1 |}
\def\ket#1{| #1\rangle}

\def\ul#1{\underline{#1}}

\def\R{\hbox{\rm I \kern-5pt R}}

\font\ticp=cmcsc10
\def\ajou#1&#2(#3){\ \sl#1\bf#2\rm(19#3)}
%
%
\lref\wiesner{S.~Wiesner, \ajou SIGACT News &15 (83) 78.} 
\lref\BBeightyfour{C. H. Bennett and G. Brassard, in {\it Proceedings of IEEE
 International Conference on Computers, Systems and Signal Processing} (IEEE,
 New York, 1984), p.~175.} 
\lref\schneier{B.~Schneier, {\it Applied Cryptography} (2nd ed., Wiley,
 New York, 1996).}
\lref\BCJL{G.~Brassard, C.~Cr\'{e}peau, R.~Jozsa and D.~Langlois, in
 {\it Proceedings of the 34th Annual IEEE Symposium on the Foundation of
 Computer Science} (IEEE Comp. Soc., Los Alamitos, California, 1993), p.~362.}
\lref\brassardcrepeau{G.~Brassard and C.~Cr\'{e}peau, 
in {\it Advances in Cryptology:
 Proceedings of Crypto'90}, Lecture Notes in Computer Science Vol 537
(Springer-Verlag, Berlin, 1991), p.~49.} 
\lref\ardehali{M.~Ardehali, quant-ph/9505019.} 
\lref\lochauprl{H.-K.~Lo and H.~Chau, \ajou Phys. Rev. Lett. &78 (97)
3410.}
\lref\mayersprl{D.~Mayers, \ajou Phys. Rev. Lett. &78 (97) 3414.}
\lref\mayerstrouble{D.~Mayers, quant-ph/9603015.} 
\lref\lochau{H.-K.~Lo and H.~Chau, \ajou Physica &D120 (98) 177.} 
\lref\lo{H.-K.~Lo, \ajou Phys. Rev. A &56 (97) 1154.} 
\lref\mayersone{D. Mayers, in {\it Proceedings of the Fourth Workshop on
 Physics and Computation} (New England Complex System Inst., Boston, 1996),
 p.~226.} 
\lref\bcms{G.~Brassard, C.~Cr\'epeau, D.~Mayers and L.~Salvail, 
quant-ph/9806031}
\lref\benoretal{M.~Ben-Or, S.~Goldwasser, J.~Kilian and A.~Widgerson, 
in {\it Proceedings of the 20th Annual ACM Symposium on the Theory
of Computing, Chicago, 1988} (Academic, San Diego, Toronto, 1991) pp. 113-131.}
\lref\blum{M.~Blum, in {\it Proceedings of the 24th IEEE Computer
Conference ( Compcon)} (IEEE, New York, 1982), pp. 133-137.}
\lref\yao{A.~Yao, in {\it Proceedings of the 26th Annual ACM Symposium
on the Theory of Computing, Montreal, 1994} (Association for 
Computing Machinery, New York, 1994), p. 67.}
\lref\kilian{J.~Kilian, in {\it Proceedings of the 20th Annual ACM Symposium
on the Theory of Computing, Chicago, 1988} (Academic, San Diego,
Toronto, 1991) p. 20.}
\lref\brassard{See e.g.~ G. Brassard, {\it Modern Cryptology: A Tutorial},
Lecture Notes in Computer Science 325 (Springer-Verlag, New York, 1988).} 

\Title{\vbox{\baselineskip12pt\hbox{ DAMTP-1997-135}\hbox{quant-ph/9810068}{}}
}{\centerline{Unconditionally Secure Bit Commitment}}{~}
\centerline{{\ticp Adrian Kent}}
\vskip.1in
\centerline{\sl Department of Applied Mathematics and
Theoretical Physics,}
\centerline{\sl University of Cambridge,}
\centerline{\sl Silver Street, Cambridge CB3 9EW, U.K.}

\bigskip

\centerline{\bf Abstract}
We describe a new classical bit commitment protocol based 
on cryptographic constraints imposed by special relativity.
The protocol is unconditionally secure against classical or quantum
attacks.  It evades the no-go results of Mayers, Lo and Chau
by requiring from Alice a sequence of communications, including
a post-revelation verification, each of which
is guaranteed to be independent of its predecessor. 
\vfill
Electronic address: apak@damtp.cam.ac.uk
\eject
\newsec{Introduction} 

The discovery of secure quantum key 
distribution\refs{\wiesner} and other applications of
quantum information has excited much interest in the general
question of precisely which cryptographic tasks can be 
guaranteed secure by physical principles.  In particular,
several papers\refs{\BBeightyfour, \brassardcrepeau, \BCJL, \ardehali,
\mayersprl, \mayerstrouble, \lochauprl, \lochau, \bcms} have 
addressed the question of whether security
can be physically guaranteed for the key crytographic primitive 
of bit commitment.

In a bit commitment protocol Alice and Bob exchange data in such a 
way that Bob obtains an encoding of a bit chosen by Alice.  
For the protocol to
be secure against Bob, it must guarantee that Bob cannot decode 
the bit until Alice chooses to reveal it by supplying further information. 
For it to be secure against Alice, it must guarantee that the 
bit is genuinely fixed between commitment and revelation: 
there must not be two different decodings of the bit which leave Alice 
free to reveal either $0$ or $1$, as she wishes. 

Bit commitment per se has obvious practical applications.
For example, a secure bit commitment protocol would allow Alice 
to make predictions which could be verified post hoc without giving
Bob any possibility of extracting information before the predicted event. 
More generally, bit commitment is a powerful cryptographic primitive.
A trusted protocol for committing a classical bit could used as a 
building block for protocols implementing
a wide range of other cryptographic tasks, including
coin tossing\refs{\blum}, 
zero-knowledge proofs\refs{\brassard}, oblivious transfer\refs{\yao}
and (hence) secure two-party computation\refs{\kilian}.   

In the standard cryptographic scenario, Alice and Bob each
occupy a laboratory.  Each trusts the integrity of their own
laboratory but nothing outside.  
It is usually implicitly assumed that the 
presumed separation of the laboratories is large compared to their
size.  In this situation, a protocol must allow for a time lapse
between the transmission of a signal and its receipt.   
However, neither party can be 
certain whether the other actually {\ it is} confined to a distant laboratory:
if it were advantageous, Bob might set up a secret laboratory adjacent
to Alice's, or vice versa.
Allowing for special relativity gives
no security advantage under these conditions, since no time lapse 
can be guaranteed, so that no arrangement of timings in a protocol
can guarantee that messages sent by Alice and by Bob were each
generated without knowledge of the other. 
Thus they are effectively restricted to protocols in which they 
sequentially exchange messages, each
waiting to receive one message before sending the next, 
and their communications may as well be taken to be 
non-relativistic.  

We refer to any bit commitment protocol
that relies on this scenario as a {\it standard} protocol. 
We refer to a protocol as classical if the protocol can be 
followed by exchanging classical information, and as quantum
if it requires the exchange of quantum information. 
We follow the formal definitions of perfect and unconditional security
given in Ref. \refs{\mayersprl}.

All standard classical bit commitment protocols are in principle insecure,
though very good practical security can be attained.  
Several quantum bit commitment schemes have been 
proposed.\refs{e.g. \BBeightyfour -- \ardehali}
But all standard quantum bit commitment schemes were also shown by 
Mayers, Lo and Chau\refs{\mayersprl, \mayerstrouble, \mayersone,
\lochauprl, \lochau} to be insecure.  
We follow general usage in referring
to the result that  unconditionally secure quantum bit commitment
is impossible as the Mayers-Lo-Chau no-go theorem or MLC theorem.  

In practice, current bit commitment protocols rely 
for their security on the assumption that some computational task 
is sufficiently hard that it cannot be carried out during the lifetime
of the protocol.  While those assumptions are generally well founded,
they never absolutely guarantee security.  
Moreover, the possible development of quantum computers 
renders the computational assumptions underlying
present day bit commitment protocols distinctly vulnerable. 
The MLC theorem tells us that
quantum technology offers no compensating solution. 
Lo has also 
shown that other two-party cryptographic tasks cannot
be securely implemented by quantum communication.\refs{\lo}

All of these no-go theorems implicitly assume that
relativity can be neglected, as is
indeed the case for standard protocols.  
Here we describe a protocol which uses a variant of 
the standard cryptographic scenario in which each party 
controls two separated sites.  Relativity plays an essential
role in this protocol: its security is
guaranteed by the impossibility of superluminal signalling.  

Variations of the standard cryptographic scenario of this type,
in which special relativity plays a r\^ole, do not seem to have been widely 
considered.  
Such protocols were, however, mentioned briefly in Mayers' 
announcement of the no-go theorem\refs{\mayersprl} for unconditionally
secure quantum bit commitment, where it is 
suggested that the no-go theorem applies
also to quantum bit commitment protocols based on special 
relativity.  

The validity of the MLC theorem in the standard 
scenario is not disputed here, but we argue for the opposite conclusion 
when special relativity
is taken into account.  We first 
describe a relativistic cryptographic scenario in which each party controls 
laboratories in two separated locations.  
These laboratories must be near to mutually agreed coordinates, 
and the protocol includes tests to verify that this is so.  
This should be stressed: neither party needs to trust the 
other's word as to the locations
of their laboratories, nor do these locations need to be declared
precisely.  

Next, we describe a bit commitment protocol in this 
scenario.  The protocol is classical: it does not require the transmission or 
processing of quantum information.  Nothing in it prevents
either party from using quantum information transmissions.  
However, the classicality of the information could be enforced by
a reasonable extra cryptographic assumption, namely the use
of channels trusted by both parties to be decohering. 
Its security can thus sensibly be analysed by considering it
either as a classical protocol or a quantum protocol. 
It is, we argue, unconditionally secure in either case.  

Ben-Or et al. (BGKW) some time ago\refs{\benoretal}  proposed an 
interesting bit 
commitment protocol which, like that presented here, depends
on separating Alice into two parties, in this case
isolated by Faraday cages.  
Its security against
quantum attacks has been discussed by Brassard et al. (BCMS) \refs{\bcms}. 
Among the significant differences between the Ben-Or et al. protocol
and the one below are that the BGKW protocol gives Bob no 
reliable test for ensuring that the two Alices are indeed
unable to communicate: unlike the present protocol, its security
is not guaranteed by physical laws.   If the isolation
is ensured by special relativity, the BGKW protocol can be seen 
as a precursor of that described here. 

The possibility of ensuring temporary isolation by special relativity
was noted by BCMS \refs{\bcms}.\foot{BCMS follow Mayers in 
concluding that unconditionally secure bit commitment is impossible.}
However, no complete discussion of the uses of relativity in 
obviating the need for trust seems to have previously appeared 
in print.  As the next section explains, the protocol given 
here uses a relativistic 
scenario in which Alice and Bob are treated symmetrically 
and in which it is demonstrably unnecessary for either party
to trust in the locations of the other.  
Finally, the key new feature of our protocol is the use of 
a sequence of communications to maintain security indefinitely.  

\newsec{Cryptography and relativity}

We now consider a cryptographic scenario in which two parties
carry out operations from separated regions in Minkowski spacetime.  
In fact, it is sufficient for the local geometry to be approximately 
Minkowski, so that the protocol can indeed be securely implemented
in the real world.  However, strictly speaking, even assuming an
approximately Minkowski background violates 
the cryptographic rule that the world outside the 
laboratory cannot be trusted.  Alice and Bob need to be confident 
that the geometry of the spacetime region is indeed nearly
flat, that they have a correct description of the local light
cones, and that there are no wormholes or other mechanisms allowing 
signalling between spacelike separated points.  

These caveats are rather irrelevant for practical applications at 
present.  It seems safe, for example, to neglect the danger that a protocol 
carried out within the solar system might be subverted by one of the parties 
surreptitiously introducing very massive bodies. 
Still, there is a theoretical case for distinguishing unconditional
security based on special relativity and on general relativity. 
We take special relativity to be the underlying theory here, and
we set $c=1$. 

Consider now the following arrangement.  Alice and Bob agree on a 
frame, on global coordinates, and on the location of two sites 
$\ul{x}_1, \ul{x}_2$. 
Alice and Bob are 
required to erect laboratories, including sending and receiving
stations, within a distance $\delta$ of the sites, where 
$ \Delta x = | \ul{x}_1 - \ul{x}_2 | \gg \delta $.  
The precise locations of the laboratories need not be disclosed: it is 
sufficient that test signals sent out from each of
Bob's laboratories receive a response within time $2 \delta$ from
Alice.   In the protocol below, Bob need not reply immediately
to Alice's communications, but the parties will probably want to test
that Bob likewise replies to Alice's test signals within time $2 \delta$
in order to confirm that the channels are working properly in both directions. 
The laboratories need not be restricted in
size or shape, except that they must not overlap.  This is implied by 
the standard assumption that Alice
and Bob are each confident of the security of their own laboratories.
We refer to the laboratories in the vicinity of $\ul{x}_i$ as 
$A_i$ and $B_i$, for $i = 1$ or $2$. 

We assume that $A_1$ and $A_2$ are collaborating with complete mutual
trust and with prearranged agreements on how to proceed, to the extent
that we identify them together simply as Alice; similarly $B_1$ and
$B_2$ are identified as Bob.  For example, considering embassies
as faithful representatives of their respective governments, we 
could take $A_1$ to be the Andorran embassy in Belize, $B_1$, 
and $B_2$ the Belizean embassy in Andorra, $A_2$.  

\newsec{A bit commitment protocol} 

We first define a classical protocol
and then examine its security against quantum attacks.   
Alice and Bob first agree a large number $N$.
For simplicity we take $N = 2^m$, where the integer $m$ is the security
parameter for the protocol.  All the arithmetic in the protocol
is carried out modulo $N$.  Before the protocol begins, 
$A_1$ and $A_2$ agree a list $\{ m_1 , m_2 , \ldots \}$
of independently chosen random numbers in the range $0, 1, \ldots , N-1$.  
The length of the list that will eventually be required is an  
exponential function of the anticipated time between commitment and 
unveiling.  Alice and Bob also fix a time interval, $\Delta t << \Delta x$, 
during which each round of communication between $A_i$ and $B_i$
(for $i=1$ or $2$) must be completed. 

The protocol now proceeds as follows.  Between time $t=0$ and
$t = \Delta t$, $B_1$ sends
$A_1$ a labelled pair $(n^1_0 , n^1_1 )$ of randomly chosen distinct
numbers in the range $0,1, \ldots , N-1$.  On receiving these 
numbers, $A_1$ returns either the number $n^1_0 + m_1$ or 
$n^1_1 + m_1$, depending whether she wants to commit a $0$ or
a $1$, quickly enough that her message ends by time 
$t =  \delta + 2 \Delta t$ and so can be received by $B_1$ before
time $2 \delta + 2 \Delta t$.  
At time $t = T = \Delta x - 2 \Delta t - 3 \delta$, $B_2$ asks $A_2$ to commit to 
him the binary form $a^1_{m-1} \ldots a^1_0 $ of $m_1$.
This is achieved by sending $A_2$ a set of $m$ labelled pairs 
$(n^2_0 , n^2_1 ), \ldots , (n^{m+1}_0 , n^{m+1}_1 )$, and 
asking $A_2$ to return $n^2_{a^1_0} + m_2 , \ldots , 
n^{m+1}_{a^1_{m-1}} + m_{m+1}$.   Bob's message is 
to be completed by time $T + \Delta t$ and Alice's by
$T +  \delta + 2 \Delta t$.   
Next, at time $t = 2 T$, 
$B_1$ asks $A_1$ to commit the binary
forms of the random numbers $m_2 , \ldots , m_{m+1}$ used 
by $A_2$.  At time $t = 3 T$,
$B_2$ asks $A_2$ to commit the
binary forms of the random numbers $m_{m+2}, \ldots , m_{m^2 + m + 1}$
used by $A_1$ in this commitment; and so forth.  These later 
exchanges are all similarly timed, so that Bob's $(N+1)$-th 
communication is completed by $N T + \Delta t$ and Alice's by
$N T + 2 \delta + 2 \Delta t$.  
The random pairs sent by the $B_i$ are all drawn from independent
uniform distributions.  

These commitments continue at regular intervals separated by 
$ T $, consuming 
increasingly long segments of the random string shared by the $A_i$, 
until one or the other of the $A_i$ --- or perhaps both, at 
spacelike separated points --- chooses to unveil the originally
committed bit.  It is assumed that the $A_i$ have previously 
agreed under which conditions either of them will unveil.  
For $A_1$ to unveil, she reveals to $B_1$ the set of random numbers used by
$A_2$ in her last set of commitments; similarly, $A_2$ unveils
by revealing to $B_2$ the random numbers last used by $A_1$.  
To check the unveiling, $B_1$ and $B_2$ send the unveiling data 
and all previous commitments to some representative of Bob.
This representative need not be in the same location as one of 
the $B_i$: if he is, only the other $B_i$ need send data.  

In any case, Bob cannot verify the unveiling at any point outside
the intersection of the future light cones of the 
points from which the $A_i$ sent 
their last communications --- i.e. the unveiling and the last 
set of commitments.  In this sense, the protocol is not complete
at the moment of unveiling: it becomes complete only when Bob 
has all the necessary data in one place.  The 
need to wait for receipt of information which is unknown to 
the unveiler $A_i$ (since it depends on the last set of pairs 
sent by $B_{3-i}$) and to the unveilee $B_i$ (since it includes
the last set of commitments sent by $A_{3-i}$) means that the 
protocol is not vulnerable to a generalised Mayers-Lo-Chau
attack.  

The protocol is clearly secure against Bob, who receives what are
to him random numbers throughout the protocol, until unveiling.  
We give here informal arguments for the insecurity against Alice.

\newsec{Security against classical attacks} 

Can Alice unveil a $0$, having committed a $1$, or vice versa? 
Note first that if $A_2$ unveils at times between
$0$ and $T$, the protocol is clearly secure.
Now suppose for definiteness that $A_1$ unveils at time between 
$ N T $ and $ (N+1) T$.  
If the $A_i$ have
followed the protocol throughout, and $A_1$ now 
gives $B_1$ the random numbers used by 
$A_2$ in her last commitment outside the future light cone of
this communication, Bob
will --- once $B_1$ and $B_2$ have had time to communicate ---
be able to decode successive commitments back through
to obtain the originally committed bit.   

On the other hand, if $A_1$ gives $B_1$ any other set of random
numbers, they will fail to correspond to a valid set of 
bit commitments with probability at least
$(1 - \frac{1}{N})$, since $A_1$ cannot yet know the pairs 
$(n^i_0 , n^i_1 )$ supplied by $B_2$ for $A_2$'s last commitment. 
So $A_1$ must supply the correct numbers.  Now if $A_2$'s last
commitment was not of the random numbers previously used by
$A_1$, a similar problem occurs.  Hence, by induction on the
total number of commitments, the protocol is secure against 
Alice.   

\newsec{Security against quantum attacks} 

Quantum attacks give Bob no advantage against an honest Alice.
His only extra freedom is to send Alice superposition states 
instead of classical descriptions of the pairs $(n^i_0 , n^i_1 )$, 
and since she can legitimately carry out measurements 
on them and follow the classical protocol, this gains him 
nothing.
We can therefore assume that Bob
sends classical signals to Alice, and that at unveiling
he carries out measurements on any superposed quantum signals sent by 
her, so as to obtain a definite set of numbers for each commitment. 

Alice's position is a little more complicated to analyse.  
Quantum theory clearly opens up new strategies for her.  
For example, following the general Mayers-Lo-Chau 
strategy\refs{\mayersprl, \lochauprl, \bcms} for cheating standard 
quantum bit commitment schemes, she can keep all her random choices 
at the quantum level.  
To do this, instead of sharing a list of random numbers from 
$0$ to $N-1$ before the protocol, $A_1$ and $A_2$ share
entangled ``quantum dice'' in correlated states of the 
form $\sum_{i=0}^{N-1} 
a_i \ket{i} \bra{i}$.   

Alice could also commit a random quantum bit --- a 
state of the form $ a \ket{0} + b \ket{1}$ --- rather than a 
fixed classical bit, and keep the committed quantum bit in 
superposition throughout, without detectably deviating from 
the protocol.  
This is no advantage if the protocol is used
for committing a prediction or some other stand-alone application.  
Alice can always commit a 
randomly chosen classical bit in any bit commitment protocol,
classical or quantum.  But it does allow Alice more general
coherent quantum attacks to be used on schemes of which
the bit commitment is a sub-protocol --- a property 
which is shared by other classical bit commitment schemes\refs{\bcms} 
and which means that classical cryptographic reductions involving
such bit commitments cannot naively be carried over into the 
quantum arena.  
 
Modulo this freedom, the informal security arguments above carry
over to the quantum case.  
Alice has no cheating strategy by which she can initially 
commit the qubit $ a \ket{0} + b \ket{1}$ and appear to follow through
the protocol for a previously agreed number of steps, while 
actually carrying out operations which give her probability greater than 
$ |a|^2 + O(1/N)$ of successfully unveiling a $0$ or greater 
than $ |b|^2 + O(1/N)$ of successfully unveiling a $1$ at the 
end of the protocol.  

\newsec{Comments} 

The protocol gives a theoretical solution to the problem of 
finding bit commitment schemes unconditionally secure over 
arbitrarily long time intervals.  As its implementation requires
channel capacity that, for a fixed separation, increases exponentially
with the commitment time, it is not a practical
solution to the problem of long term bit commitment.  
For example, taking the security parameter $m=10$ and the separation 
$\Delta x = 0.1 {~\rm sec}$, and assuming $100$ gigabaud channels, 
the number of rounds of iterated
commitments presently practical is roughly $10$.

For the moment, though, we see the protocol's main interest as an existence
theorem.  It demonstrates that taking special relativity into account
changes the cryptographic security attainable through information 
exchanges, and it shows that the interplay 
between special relativity, cryptographic security and 
channel capacity is a fertile area for investigation. 

The reason relativity helps is simple.  In effect, it allows Alice and Bob
to construct a communication channel with a time delay which they
can both trust, despite their mistrust of the world outside their
laboratories.   Any trusted time delayed channel allows temporary
bit commitment, and the above protocol demonstrates that indefinite
bit commitment can then be achieved by recursively iterating bit
commitments across the channel.  

It is worth noting that trusted,
although not perfectly secure, time delay could also be enforced by 
physical means.  Alice and Bob could, for example, watch carrier
pigeons going between their laboratories.  It could
also be enforced by a sequence of computational bounds.  
Suppose, for example, that
Alice and Bob can always
be confident of keeping abreast of technological developments, in
the sense that at any given time they can find a computational task
which they are confident cannot be solved within $1$ time unit. 
They can then, for as long as their channel capacity permits, 
use the iteration strategy above to achieve 
indefinitely secure bit commitment from a sequence of 
standard classical bit 
commitment protocols which use their temporarily secure bounds
and are secure against the receiver.  
This may, in fact, be a more immediately practical application.  

\vskip15pt
\leftline{\bf Acknowledgments}

I am very grateful to Ross Anderson, Charles Bennett, 
Andras Bodor,  Gilles Brassard, Jeff Bub, H.F. Chau, Claude Cr\'epeau, 
Peter Goddard, Hoi-Kwong Lo, Dominic Mayers, Roger Needham, 
Sandu Popescu, Louis Salvail and John Smolin for extremely 
helpful discussions and advice.  
I thank the Royal Society for financial support.  
\listrefs
\end